\documentclass[12pt]{article}
\usepackage{latexsym}
\usepackage{graphicx}
\usepackage{indentfirst}
\begin{document}

\begin{center}
{\bf \LARGE  Differential Cross Sections for  \\ \vspace*{.4cm} Electromagnetic Dissociation}
\end{center}

\begin{center}
John W. Norbury $^{\rm a, b, \star}$,  Anne Adamczyk $^{\rm a}$ \\
\end{center}

\begin{center}
{\small  {\em $^{\rm a}$ Worcester Polytechnic Institute, Worcester, MA, 01609, USA

$^{\rm b}$ University of Southern Mississippi, Hattiesburg, MS, 39406, USA}}
 \end{center}

\begin{center}
 {\em email:} john.norbury@usm.edu, 
anniea@wpi.edu
\end{center}

\noindent \hrulefill

\noindent {\bf Abstract}\\

Differential cross sections for electromagnetic dissociation in nucleus-nucleus collisions are calculated.  The kinetic energy distribution is parameterized with a Boltzmann distribution and the angular distribution is assumed isotropic in the projectile frame.  In order to be useful for three-dimensional transport codes,  these cross sections are available in both the projectile and lab frames.  Comparison  between theory and experiment is good.
The formalism  applies to single and multiple nucleon removal, $\alpha$ particle removal,  and fission in electromagnetic reactions of nuclei.\\



\noindent \hrulefill

\section{Introduction}

Nucleus-nucleus collisions can be mediated by either the strong or electromagnetic (EM) force. A reaction proceeding via the EM force is often called  {\em Electromagnetic Dissociation} (EMD).
There have been many studies of single \cite{charge, single, quad2, quad1, reanalysis, marsha, explanation2, explanation1}  and double \cite{2neutron} nucleon removal  via EMD. Fission processes have also been studied  \cite{fission, wheeler}.
For single and few nucleon removal, EMD cross sections can be just as large or larger than strong interaction cross sections \cite{acta}.  Pair production via EM processes can also be significant \cite{pair}.  Transport codes such as HZETRN \cite{develop} and FLUKA \cite{fluka} include the calculation of total EMD cross sections by using currently available parameterizations of the total cross sections \cite{apj2, apj1, nimb}. Differential EMD cross sections are necessary because fully three-dimensional transport codes require energy and angular differential cross sections.   These  cross sections are the subject of the present paper.

To improve run-time, transport codes often require parameterizations of cross sections. Parameterizations of total cross sections are available  \cite{apj2, apj1, nimb}. The differential cross sections developed in the present work are written in terms of total cross sections. If the parameterized form of the total cross sections  \cite{apj2, apj1, nimb} is used, then the differential cross sections are also parameterized. 

Even though the discussion will be presented in terms of single nucleon removal, the results also apply to multiple nucleon removal and
 $\alpha$ particle removal, by changing the branching ratios. Since fission usually involves three or more fission products, all of the formalism for the kinematics presented also applies to fission. This requires using the total fission cross sections \cite{fission, wheeler} as input to the differential cross sections. Using appropriate branching ratios and fission decay product masses will result in differential fission cross sections for a specified fission product.

\section{Photonuclear cross sections}

   A vital component to any EMD cross section calculation is the related photonuclear cross section, which is presented in this section.  Included is a discussion of the differential photonuclear cross section that will be important when deducing the nucleus-nucleus differential  EMD cross section. We also present a reminder of the total cross section parameterization, because this will be used in obtaining differential cross sections.

\subsection{Angular Distribution}

In the simplest compound nucleus model, the photonuclear angular distribution is 
approximately isotropic \cite{preston, satchler, paul, handbook},   i.e.
\begin{eqnarray}
\frac{d\sigma_{\gamma A}(E_\gamma)}{d\Omega_N} = K  \label{iso1}
\end{eqnarray}
where $K$ is a constant with respect to angle (but of course it will depend on $E_\gamma$). It is trivial to evaluate $K$ from the total cross section because
\begin{eqnarray}
\sigma_{\gamma A}(E_\gamma) = \int \frac{d\sigma_{\gamma A}(E_\gamma)}{d\Omega_N} \;d\Omega_N = 4\pi K .
\label{iso2}
\end{eqnarray}
Now an isotropic angular distribution can be written as
\begin{eqnarray}
\frac{d\sigma_{\gamma A}(E_\gamma)}{d\Omega_N} = \frac{\sigma_{\gamma A}(E_\gamma) }{4\pi}_.  \label{photoangular}
\end{eqnarray}

\subsection{Spectral Distribution}

The energy level density can be approximated   \cite{frobrich},  \cite{feshbach} (pp.  326)
 by a Boltzmann distribution
$
\rho(E) \sim  e^{-E/k\Theta}
$
with the nuclear temperature given by
\begin{eqnarray}
k\Theta = \sqrt{ \frac{10 \; E_\gamma }{A_P}}  \label{nucltemp}
 \end{eqnarray}
where $\Theta$ is the nuclear temperature and $k$ is the Boltzmann constant. Therefore, the photonuclear spectral distribution is written as
\begin{eqnarray}
 \frac{d\sigma_{\gamma A} (E_\gamma)}{dE_N} =C\; T_N\; e^{-T_N/k\Theta}  \label{photospectral-1}
\end{eqnarray}
where $T_N$ is the kinetic energy of the emitted nucleon.
The constant $C$ is determined by the
requirement 
\begin{eqnarray}
\sigma_{\rm tot}(E_\gamma) = \int dE_N \;\frac{d\sigma_{\gamma A} (E_\gamma)}{dE_N} 
\end{eqnarray}
where $\sigma_{\rm tot}(E_\gamma)$ is the photonuclear total cross section.
For simplicty, we shall first assume that the limits of integration are $0$ and $\infty$. Then
\begin{eqnarray}
\sigma_{\rm tot}(E_\gamma) = \int_0^\infty dE_N \;\frac{d\sigma_{\gamma A} (E_\gamma)}{dE_N}  
=C\int_0^\infty dT_N \;  T_N e^{-T_N/k\Theta} 
= C (k\Theta)^2
\end{eqnarray}
giving
\begin{eqnarray}
 \frac{d\sigma_{\gamma A} (E_\gamma)}{dE_N} =\frac{\sigma_{\rm tot}(E_\gamma)}{(k\Theta)^2}\; T_N \;e^{-T_N/k\Theta}
 \label{finalphotonuclspectral}.
\end{eqnarray}
 The more  general calculation with arbitrary limits, $T_{\rm min}$ and $T_{\rm max}$, gives
\begin{eqnarray}
 \frac{d\sigma}{dE_N} &=&\frac{\sigma_{\rm tot}(E_\gamma)}{ k\Theta(T_{\rm min}+k\Theta)
 e^{-T_{\rm min}/k\Theta} -  k\Theta(T_{\rm max}+k\Theta) e^{-T_{\rm max}/k\Theta}} \; T_N \;e^{-T_N/k\Theta} \\
\nonumber \\
&=&\frac{\sigma_{\rm tot}(E_\gamma)}{ k\Theta(T_{\rm min}+k\Theta)
 e^{-T_{\rm min}/k\Theta} } \; T_N \;e^{-T_N/k\Theta}  \hspace*{1.5cm} {\rm for} \;\; T_{\rm max}=\infty.
\end{eqnarray}
This reduces to the above result, equation (\ref{finalphotonuclspectral}), when $T_{\rm min}=0$ and $T_{\rm max}=\infty$.

\subsection{Double Differential Cross Section}
From reference \cite{handbook} (pp.  27, 40) (with $f=0$), the photonuclear double differential cross section can be expressed as
\begin{eqnarray}
\frac{d^2\sigma_{\gamma A} (E_\gamma)}{dE_N d\Omega_N} = \frac{1}{4\pi} \frac{d\sigma_{\gamma A} (E_\gamma)}{dE_n}  \label{photodouble}
\end{eqnarray}
which corresponds to an isotropic angular distribution.

\subsection{Total Cross Section}

The  above equations for the photonuclear differential cross sections (\ref{photoangular}),(\ref{finalphotonuclspectral}), and (\ref{photodouble}) were all written in terms of the photonuclear total cross section. We now present relevant equations to calculate the photonuclear total cross section.
 The photonuclear total cross section for producing particle X  is  \cite{norbury}
 \begin{eqnarray}
 \sigma(E_\gamma,X) = g_X \sigma_{\rm abs} (E_\gamma)  \label{branch}
 \end{eqnarray}
where $g_X$ is the branching ratio and $\sigma_{\rm abs} (E_\gamma)$ is the photonuclear absorption cross section, which is parameterized as
\begin{eqnarray}
 \sigma_{\rm abs} (E_\gamma)  = \frac{\sigma_m}{1 + \left [ (E_\gamma^2 - E_{\rm GDR}^2)^2/E_\gamma^2 \Gamma^2  \right ]}_.
 \end{eqnarray}
Here, $E_{\rm GDR}$ is the energy at which the photonuclear cross section has its peak value and $\Gamma$ is the width of the electric dipole (E1) giant dipole resonance. Also,
$
\sigma_m = \frac{\sigma_{\rm TRK}}{\pi \Gamma/2}
$
 with the Thomas-Reiche-Kuhn cross section given by \cite{norbury}
$
\sigma_{\rm TRK} = \frac{60 N_P Z_P}{A_P}\; {\rm MeV} \; {\rm mb}
$
with the subscript P referring to excitation of the projectile.  The GDR energy is
 \begin{eqnarray}
E_{\rm GDR} =  \frac{\hbar c}{\left [\frac{m^* c^2 R_0^2}{8J}(1+u -\frac{1+\epsilon +3u}{1+\epsilon +u}\, \epsilon)\right]^{1/2}}
 \end{eqnarray}
 with
$
u = \frac{3J}{Q^\prime} A_P^{-1/3}
$
 and
 $
R_0=r_0 A_P^{1/3}
$
The parameters are
 $
\epsilon = 0.0768, 
Q^\prime = 17 \; {\rm MeV}, 
J =  36.8 \; {\rm MeV}, 
r_0 = 1.18 \; {\rm fm},$ and  
$ m^* = 0.7  \;  m_{\rm nucleon} 
$.

\section{Nucleus-Nucleus  Cross Sections}
In this section we shall be evaluating total and differential cross sections. Total cross sections are, of course, Lorentz invariant. All differential cross sections are evaluated in the 
 rest frame of the compound nucleus. If the
compound nucleus is the projectile, then this must be transformed to the Lab frame for use 
in transport codes. No transformation is necessary if the compound
nucleus is the target, because the target is at rest in the Lab.

\subsection{Total Cross Section}
The  total  absorption cross section for electromagnetic nucleus-nucleus reactions  
can be written in the form
\begin{eqnarray}
\sigma_{AA} = \int dE_\gamma \;N(E_\gamma) \; \sigma_{\gamma A} (E_\gamma)    \label{aatotal}
\end{eqnarray}
where $N(E_\gamma)$ is the Weiszacker-Williams virtual photon spectrum  and 
 $\sigma_{\gamma A}(E_\gamma)$ is the
photonuclear total cross section. The total cross section for producing a particle $X$ is again just the total absorption cross section multiplied by the branching ratio
\begin{eqnarray}
\sigma_{AA}(X) = \int dE_\gamma \;N(E_\gamma) \; \sigma_{\gamma A} (E_\gamma, X)    \label{aatotal}
\end{eqnarray}
where $\sigma_{\gamma A} (E_\gamma, X)   \equiv \sigma (E_\gamma, X)   $ from equation (\ref{branch}) and $\sigma_{AA}(X)  = g_X \sigma_{AA}$.  All  the differential and total cross sections that follow can be written like the two equations  above. That is, they  can either refer to absorption cross section or cross section for producing a particle $X$.

\subsection{Angular Distribution}
The spectator nucleus is nothing more than a source of virtual photons, therefore the angular
 and spectral distributions may also be
written in the form of equation (\ref{aatotal}). Thus 
 the angular  distribution, for emission of a nucleon $N$ in the direction $\Omega_N$,  is
 \begin{eqnarray}
\frac{d\sigma_{AA}}{d\Omega_N} = 
\int dE_\gamma \;N(E_\gamma) \;\frac{d\sigma_{\gamma A}(E_\gamma)}{d\Omega_N}    \label{aaangular}
\end{eqnarray}
where $\frac{d\sigma_{\gamma A}(E_\gamma)}{d\Omega_N}$ is the photonuclear angular distribution 
 for emission of a nucleon $N$ in the direction $\Omega_N$.
If the photonuclear angular distribution is approximately isotropic, then the use of equations
(\ref{iso1}-\ref{photoangular}) and (\ref{aaangular}) gives
\begin{eqnarray}
\frac{d\sigma_{AA}}{d\Omega_N} = \frac{\sigma_{AA}}{4\pi} _.
\label{aaangular2}
\end{eqnarray}

\subsection{Spectral Distribution}
The spectral distribution, for emission of a nucleon $N$ with energy $E_N$, may also be
written in the form of equation (\ref{aatotal}), namely
\begin{eqnarray}
\frac{d\sigma_{AA}}{dE_N} = \int dE_\gamma \;N(E_\gamma) \;\frac{d\sigma_{\gamma A}(E_\gamma)}{dE_N} 
   \label{aaspectral}
\end{eqnarray}
where $\frac{d\sigma_{\gamma A}(E_\gamma)}{dE_N}$ is the photonuclear spectral distribution 
 for emission of a nucleon $N$  with energy $E_N$.  Note that the spectral distribution cannot be taken outside the integral because the nuclear temperature $\Theta$ depends on the photon excitation energy  $E_\gamma$.

\subsection{Double differential Cross Section}
The double differential  cross section, for emission of a nucleon $N$ with energy $E_N$ in the 
direction $\Omega_N$, may also be
written in the form of equation (\ref{aatotal}), namely
\begin{eqnarray}
\frac{d^2\sigma_{AA}}{d\Omega_N dE_N} = \int dE_\gamma \;N(E_\gamma) \;\frac{d^2\sigma_{\gamma A}(E_\gamma)}{d\Omega_N dE_N}   
\label{aadouble}
\end{eqnarray}
where $\frac{d^2\sigma_{\gamma A}(E_\gamma)}{d\Omega_N dE_N}$ is the photonuclear double differential cross section
 for emission of a nucleon $N$  with energy $E_N$ in the direction $\Omega_N$.
 If the photonuclear angular distribution is isotropic, we use (\ref{photodouble}) to give
\begin{eqnarray}
\frac{d^2\sigma_{AA}}{d\Omega_N dE_N} &=&  \frac{1}{4\pi} \int dE_\gamma \;N(E_\gamma) \;\frac{d\sigma_{\gamma A}(E_\gamma)}{dE_N}   \\
&=&   \frac{1}{4\pi} \frac{d\sigma_{AA}}{dE_N} 
\end{eqnarray}
which is exactly analagous to (\ref{photodouble}).

\section{Results}

 The nucleus-nucleus differential cross sections (\ref{aaangular}), (\ref{aaspectral}), and (\ref{aadouble}) all  
involve a photonuclear differential cross section or a total cross section. The photonuclear differential cross
sections are all evaluated in the rest frame of the nucleus undergoing the photonuclear reaction.
All differential cross sections in a  radiation transport code are required in the Lab frame.  If the projectile nucleus is
the one undergoing photodisintegration, then the nucleus-nucleus differential cross sections 
(\ref{aaangular}), (\ref{aaspectral}), and (\ref{aadouble}) are first evaluated in the projectile frame. They must then be Lorentz transformed to the Lab frame. The technique for doing this is described in reference \cite{joachain}

The nucleus-nucleus double differential cross sections are presented in Figs.  1 and 2. 
 The cross sections are for the reaction
\begin{eqnarray}
{\rm {^{28}Si} + {^{208}Pb} \rightarrow n +  {^{27}Si} + ^{208}Pb }
\end{eqnarray}
at 14.6 A GeV.
These are all obtained simply by taking the corresponding photonuclear cross section and integrating over the virtual photon spectrum. The double differential cross section in the lab frame shown in Fig.  2 shows the double peak feature discussed by Hagedorn \cite{hagedorn}. This double peak comes about because there is a single peak in the spectral distribution in the projectile frame. This single peak in the spectral distribution in the projectile frame gets boosted both forward and backward in the lab frame, depending on the kinematic conditions.  
The single peak in the projectile frame shown in Fig. 1, contains contributions from particles which move both forward and backward in the projectile frame. For example,  a particle near the peak with a projectile kinetic energy of 3 MeV can be either moving forward or backward in the projectile. Fig. 1 can only show the 3 MeV energy and cannot show whether it moves forward or backward.
However after Lorentz transformation to the lab frame, and depending on the particle energy \cite{hagedorn},  the forward moving particles receive a boost from the projectile frame velocity whereas the backward moving particles end up with a different speed in the lab. Thus the single peak which appeared in the projectile frame can appear as a double peak in the lab frame \cite{hagedorn}
 The high energy of the projectile  also causes angles in the lab frame to be strongly forward peaked.\\

  \hspace*{1cm}   \includegraphics[width=12cm]   {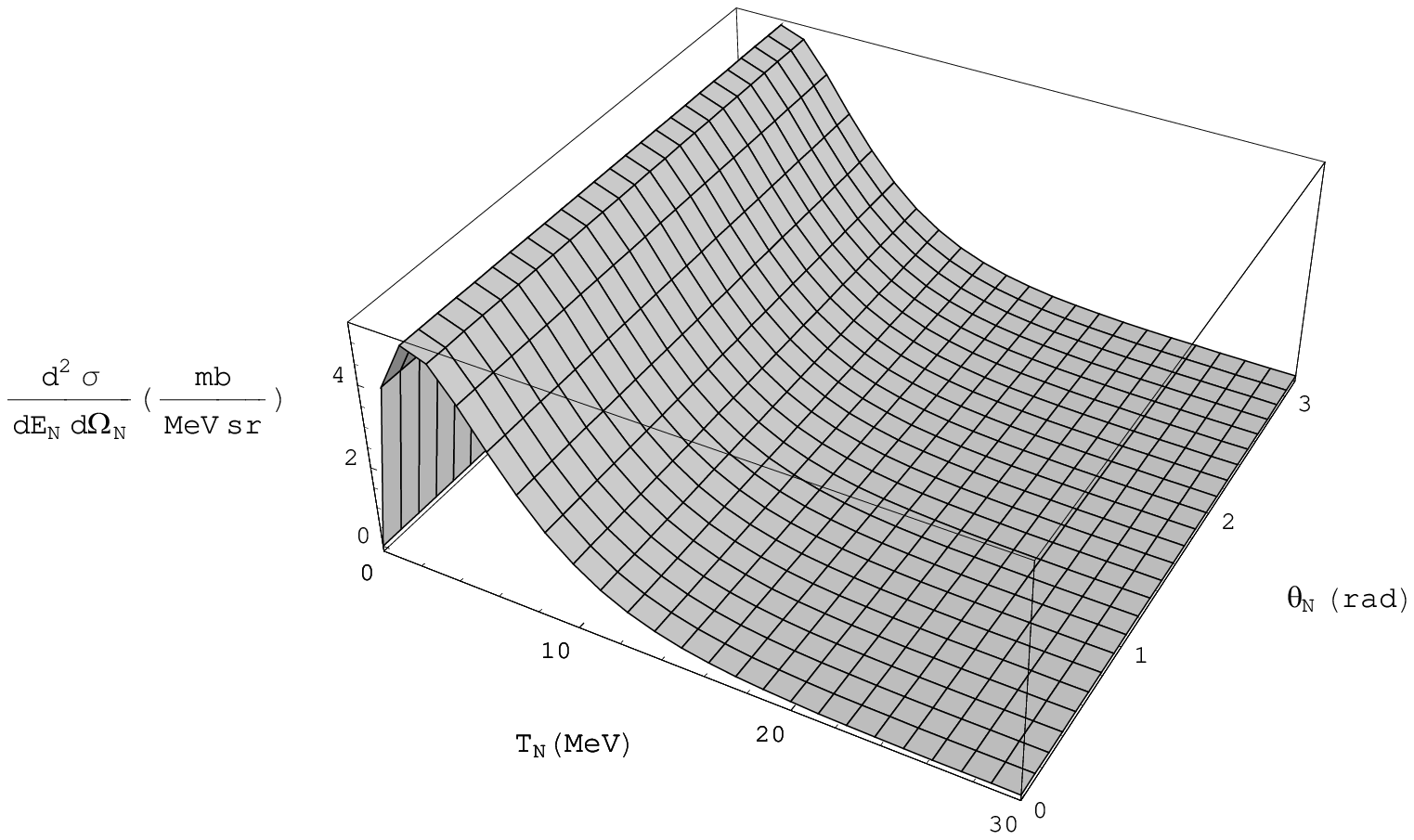} 
  \begin{quote}
Fig.  1.   Nucleus-nucleus   double differential cross section in Projectile  frame. 
\end{quote}

  \hspace*{1cm}   \includegraphics[width=12cm]   {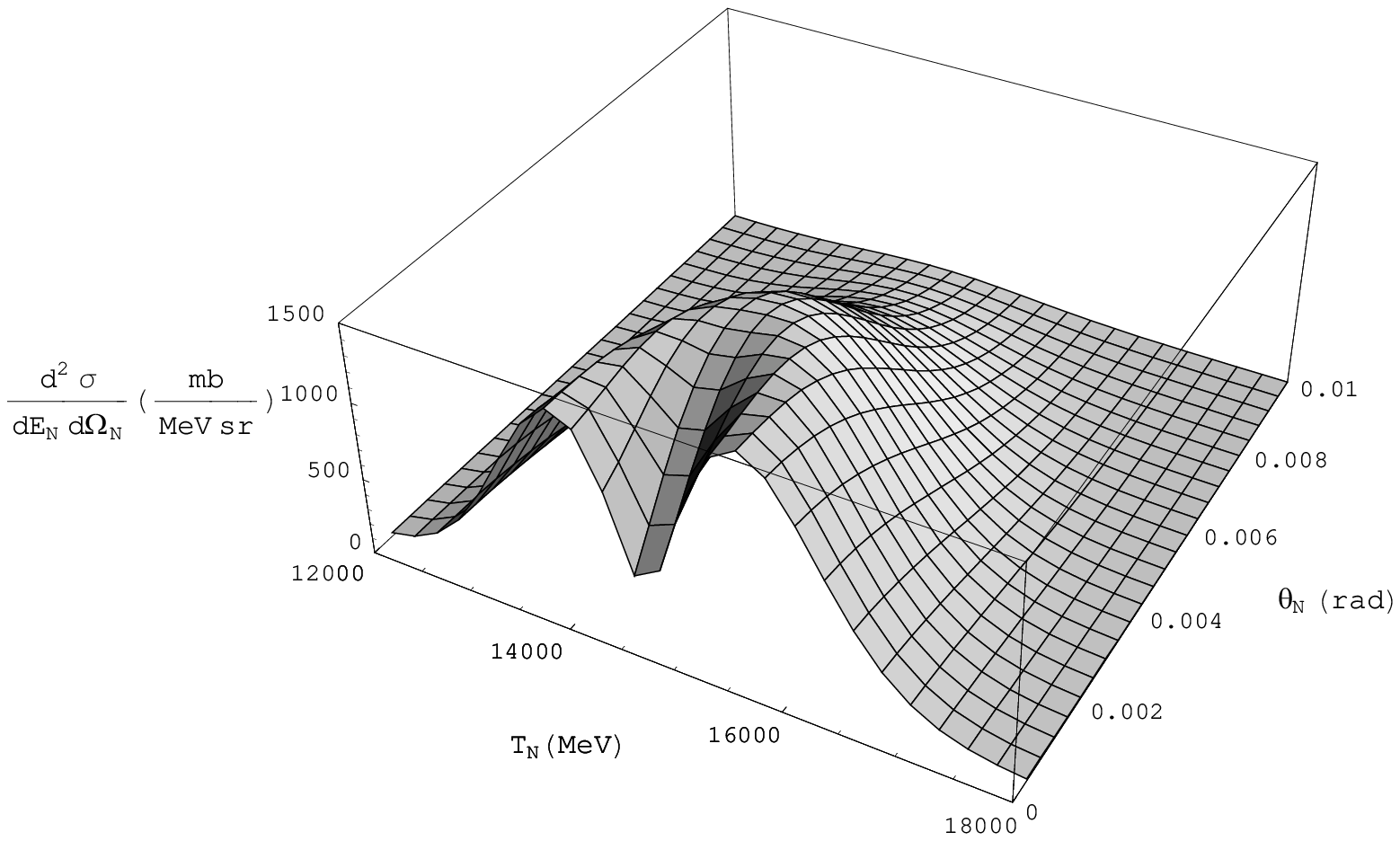} 
  \begin{quote}
Fig.  2.   Nucleus-nucleus   double differential cross section in Lab frame. 
\end{quote}

\subsection{Comparison to Experiment }

There is very little experimental data concerning differential cross sections for electromagnetic dissociation. The best data available has been measured by Barrette et al. \cite{barrette92, barrette95}, but much of their data involved spectral distributions of excitation energy.   However, a notable feature of their measurements is that  all of their angular distributions are approximately isotropic in the projectile frame, which agrees  with the assumption of the present work. See equation (\ref{aaangular2}).
Some kinetic energy distributions have been measured for outgoing neutrons and protons. See Fig.  13 of Barrette et al \cite{barrette95}. In Figs. 3 and 4, we have compared our theory to this data.
The experimental work quoted arbitrary units, so that it was necessary to fit the absolute value (peak cross section) to the experiment.  The best fit is obtained by modifying equation (\ref{nucltemp}) to
\begin{eqnarray}
k\Theta = \sqrt{ \frac{20 \; E_\gamma }{A_P}}_. \label{nucltemp2}
 \end{eqnarray}
 The comparison between theory and experiment is good. A better fit can probably be obtained with a more sophisticated approach to calculating the nuclear temperature and the spectral distribution (\ref{finalphotonuclspectral}).\\\\

  \hspace*{1cm}   \includegraphics[width=12cm]   {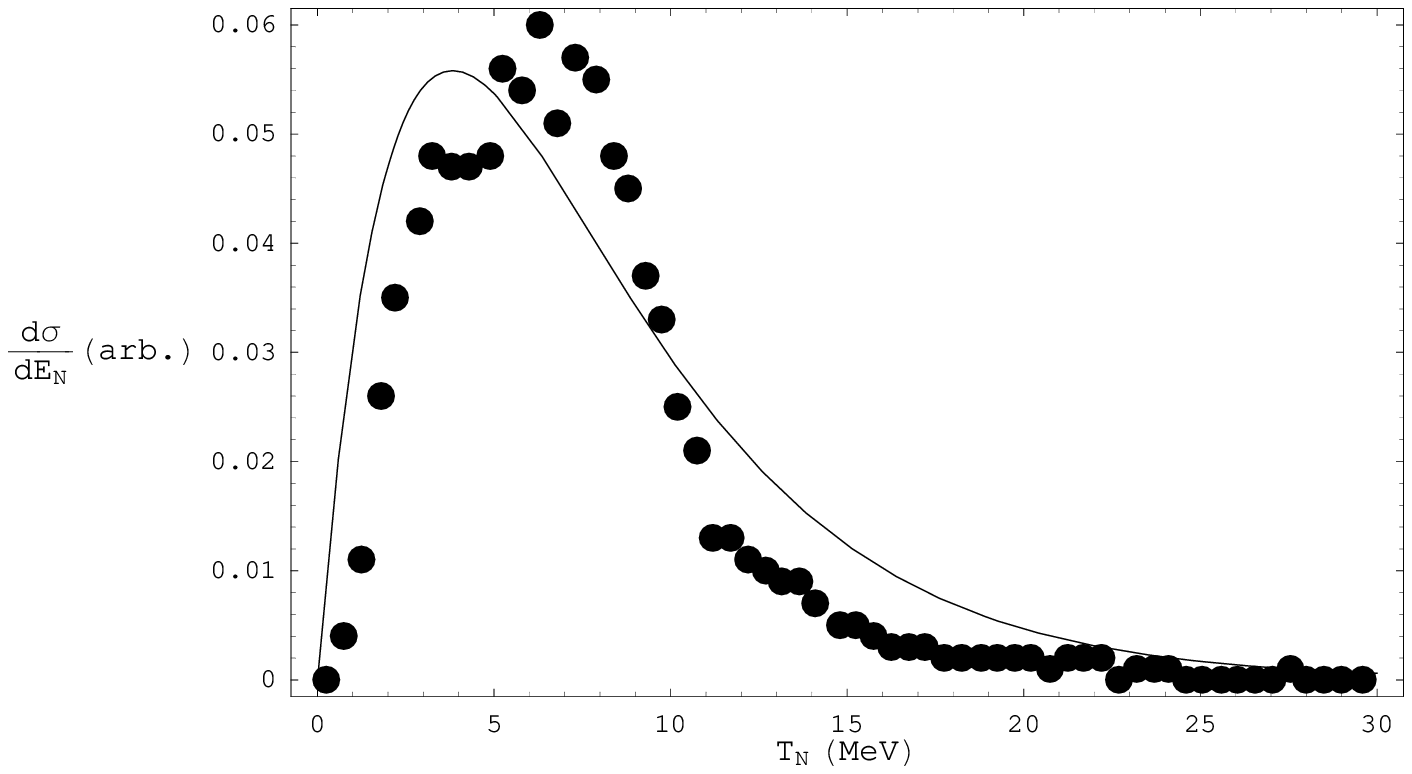}
  \begin{quote}
Fig.  3.   Comparison between theory and experiment for proton kinetic energy distribution in the projectile frame. The reaction is 
$^{28}Si +Pb \rightarrow 1p +   {^{27}Al} +Pb$ at 14.6 A GeV. Cross section units are arbitrary. Experimental data is from  Fig.  13 (b) of reference \cite{barrette95}. Error bars are smaller than symbol sizes.
 \end{quote}

  \hspace*{1cm}   \includegraphics[width=12cm]   {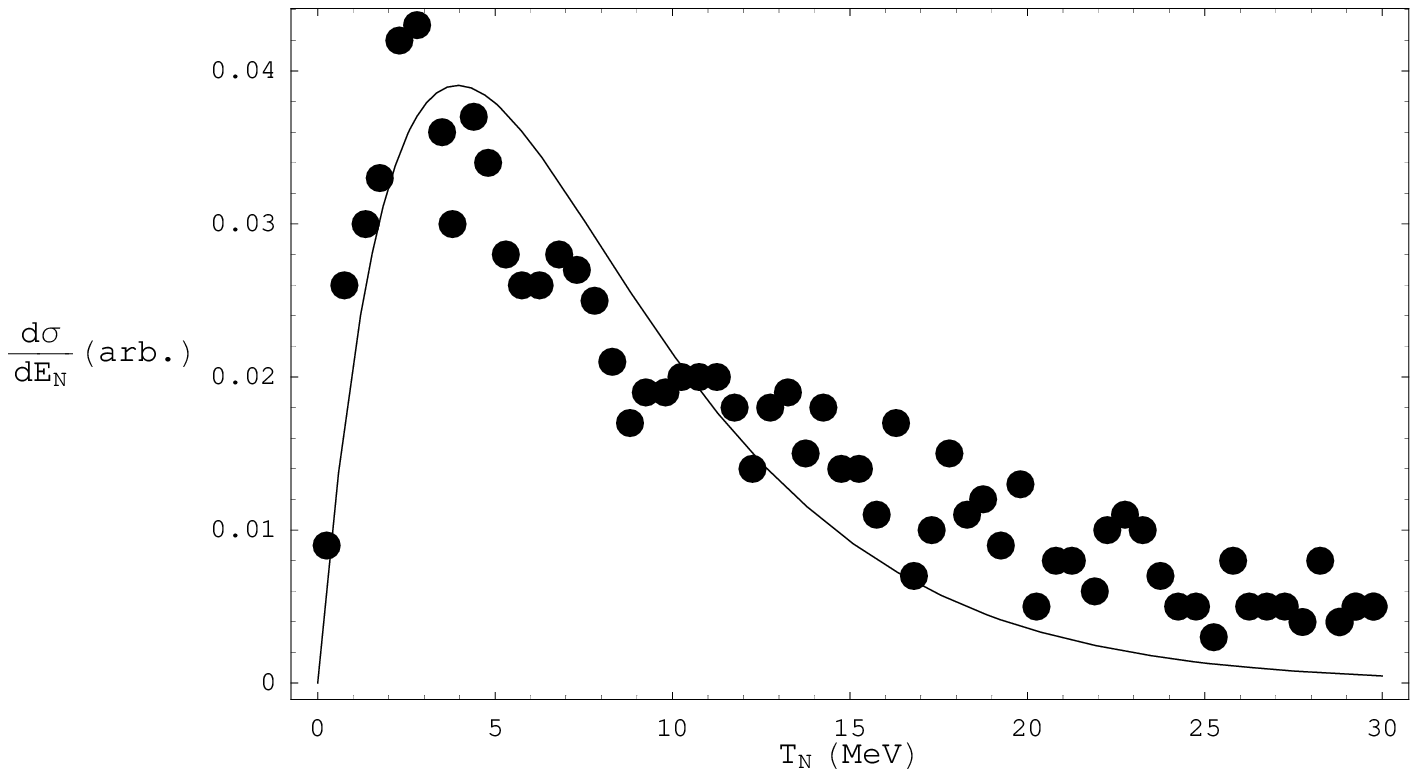}
  \begin{quote}
Fig.  4.    Comparison between theory and experiment for neutron kinetic energy distribution in the projectile frame. The reaction is  $^{28}Si +Pb \rightarrow 1n + {^{27}Si}  +Pb$ at 14.6 A GeV. Experimental data is from  Fig.  13 (c) of reference \cite{barrette95}.  Cross section units are arbitrary.  Error bars are smaller than symbol sizes.

 \end{quote}
 
 \subsection{Single and Multiple Nucleon, $\alpha$ particle removal, and fission}
 
All the above examples have focused on single nucleon removal. However, this formalism also applies to multiple nucleon removal, $\alpha$ particle removal, and fission. For multiple nucleon and $\alpha$ particle removal, just change the value of the branching ratio. When $\alpha$ particles are produced, simply use the $\alpha$ particle mass in the differential cross section formulas.
When nuclei undergo fission, there are always  three or more particles produced as fission products \cite{krane}. Therefore, all of the relativistic kinematics formalism developed above also applies to fission. The total cross section is given by that developed in reference \cite{fission}. To find the total cross sections for a particular fission fragment just use its branching ratio. The differential cross sections for a particular fission fragment is obtained by inserting the fragment mass of interest.

\section{Conclusions}
This paper presents calculations of differential cross sections for electromagnetic dissociation in nucleus-nucleus collisions. The kinetic energy distribution is parameterized with a Boltzmann distribution and the angular distribution is assumed isotropic in the projectile frame.  The results are presented in a form in which they can be immediately used in fully three-dimensional   transport codes that require differential cross sections in the lab frame. Cross sections are isotropic in the projectile frame and are in agreement with experiment.  Spectral distributions in the projectile frame are compared to experimental results and are found to be in good agreement. 
Using parameterized total cross sections as input \cite{apj2, apj1, nimb} yields parameterized differential cross sections.
The results are applicable to single and multiple nucleon removal, $\alpha$ particle removal, and electromagnetic fission.

\begin{center}
{\bf Acknowledgements}
\end{center}

\noindent
This work was supported by NASA grants  NNL05AA05G and NNL05AA06H. JWN wishes to thank Dr. Stepan Mashnik for useful correspondence.

\end{document}